\documentclass[pre,a4paper,twocolumn]{revtex4}
\usepackage{epsfig}
\usepackage{amsmath}
\usepackage{psfrag}

\newcommand{\bq}{\begin{equation}}
\newcommand{\eq}{\end{equation}}
\newcommand{\ba}{\begin{eqnarray}}
\newcommand{\ea}{\end{eqnarray}}

\begin{document}
\title{\bf Finite size effects in Barab\'{a}si-Albert growing networks}
\author{B. Waclaw}\thanks{Email address: 
bwaclaw@th.if.uj.edu.pl} 
\author{I. M. Sokolov}

\affiliation{Marian Smoluchowski Institute of Physics, 
Jagellonian University, Reymonta 4, 30-059 Krak\'ow, Poland, \\
Institut f\"{u}r Physik, Humboldt Universit\"{a}t zu Berlin, Newtonstrasse 15, 12489 Berlin, Germany}

\begin{abstract}
We investigate the influence of the network's size on the degree distribution $\pi_k$ in 
Barab\'{a}si-Albert model of growing network with initial attractiveness. Our approach based on 
 moments of $\pi_k$ allows to treat analytically several variants of the model and to calculate the 
cut-off function giving finite size corrections to $\pi_k$. We study the effect of initial 
configuration as well as of addition of more than one link per time step.
The results indicate that asymptotic properties of the cut-off depend only on the exponent 
$\gamma$ in the power-law describing the tail of the degree distribution. The method presented is this 
paper is very general and can be applied to other growing networks.
\end{abstract}

\maketitle

\section{Introduction}
Complex networks have been widely studied by physicists and researchers in many other fields. 
It was pointed out that many natural and social networks are scale-free (SF), i.e. 
that the degree distribution $\pi_k$ giving the probability that randomly chosen node has exactly 
$k$ nearest neighbors behaves as $\sim k^{-\gamma}$ for sufficiently large degrees 
$k$, typically with the exponent $\gamma$ laying between $2$ and $4$ \cite{cn1,cn2}. 
This most striking feature of real-world complex networks is well described within 
theoretical models proposed during the last decade. However, in any finite network the power-law 
behavior of the degree distribution $\pi_k$ can hold only for the values of $k$ small enough 
in comparison to the number of nodes $N$. Both experimental data and theoretical models of 
scale-free networks indicate that the behavior of $\pi_k$ for $k \gg 1$ for a finite network 
exhibits two regimes: below some $k_{max}$ it follows the power-law behavior as in an infinite 
network while above $k_{max}$ it displays a much faster decay. 
The characteristic degree $k_{max}$ which separates these two regimes is called a cut-off. 
Intuitively, the cut-off is due to the fact that since the overall number of links present 
in a finite, non-degenerated graph is bounded from above, so is also the degree of each node. 
The truncation of the power-law affects many properties of networks, especially the 
percolation ones, important in models for infection propagation of real diseases or computer viruses.
We note here that the cut-offs due to finite size effects have to be distinguished from the 
distributions' changes when a small sample of the network is considered, as discussed in 
Ref. \cite{Stumpf}, the fact which makes the analysis of real networks very involved.

Many attempts were undertaken to estimate the position of the cut-off  for different network models.
Here we consider sparse networks with fixed average degree (which corresponds to $\gamma>2$), 
and focus on the cut-off stemming from finite-size effects. They depend strongly on 
whether the network is growing and thus exhibits correlations age-degree, or is a 
homogeneous one \cite{homog}, obtained in a sort of 
thermalization process leading to statistical equivalence of all nodes. The results for 
homogeneous networks  \cite{extr1,extr2,bk} indicate that $k_{max}$ should scale as $\sim N^{1/(\gamma-1)}$ 
for $\gamma \ge 3$ and as $\sim N^{1/2}$ for $2 < \gamma < 3$ as a result of structural constraints
leading also to the occurrence of correlations between node degrees.
These results based either on probabilistic arguments or on extreme values statistics did not provide 
an explicit form of the degree distribution for a finite network. On the other hand, the 
statistical ensemble approach of Ref. \cite{dr3} lead to $k_{max}$ scaling rather like 
$N^{1/(5-\gamma)}$ for $2<\gamma \leq 3$, at variance with the prior results.

The models of growing networks allow for a more rigorous treatment. In \cite{kr2} a simple model 
of growing tree network has been solved exactly. The authors have deduced the form of corrections 
to the degree distribution and found that the cut-off scales as $\sim N^{1/(\gamma-1)}$ for any exponent 
$\gamma$ larger than two. The scaling for $\gamma<3$ is thus different from that observed for homogeneous 
networks and can be understood as a result of degree-degree correlations \cite{extr2}.

In the present paper we give a general approach to obtaining degree distributions in growing networks 
of finite size. It is based on moments of connectivity distribution $\pi_k$ and works for any case 
for which one is able to write the rate equation for $\pi_k$ depending on the size $N$. Because of 
its generality, the method is interesting by itself.
In the paper we discuss its application to the Barab\'{a}si-Albert (BA) model with initial 
attractiveness $a_0$, introduced in different contexts several times in the past \cite{willis}. 
The model is defined as follows. Starting from a finite graph of $n_0$ nodes, at each step a new node 
is introduced and joined to $m$ previously existing nodes with probability being proportional to $k+a_0$, 
where $k$ is the degree of a node to which new link is established. This rule, called preferential 
attachment, causes that nodes with high degrees get connected to more and more nodes. 
The simplest version of that model with $a_0=0$ was discussed by Barab\'{a}si and Albert \cite{bamodel}. 
An effective algorithm of building such a network with the attachment probability proportional to $k$ 
consists in focusing on the links, and not on the nodes of the graph. Each link $ij$ is then considered 
as a couple of oriented links $i\to j$ and $j\to i$ pointing into opposite directions. Because the 
probability of picking at random an oriented link pointing onto the node with degree $k$ is 
proportional to $k$, the preferential attachment rule is simply realized by choosing that node 
as the one to which the newly introduced node is attached.

The model has been extended to the case of $a_0\neq 0$ and solved in the thermodynamical 
limit in \cite{dr2}. The popularity of this model is partially explained by its three important properties: 
i) the generated network is scale-free with tunable exponent $\gamma=a_0+3$, 
ii) the generated network is a small world one, iii) the algorithm can be easily implemented on a computer.
The preferential attachment rule can be generalized to non-linear attachment kernels \cite{kr,kr3}.
These however do not give pure scale-free behavior and thus will not be discussed here. 

Some of the results we present here were obtained previously in Ref. \cite{kr2}.
Our approach is however different from the one of \cite{kr2} and allows for solving 
more sophisticated variants of the model. The method is explained in Section II where it is introduced
on the example of the simplest BA model with $m=1$ and $a_0=0$. In Section III we consider the 
influence of the initial configuration on the cut-off function.
This is of particular importance since it has been pointed out \cite{kr2,dr1} that 
the functional form of the cut-off depends strongly on the seed graph. We show however that the 
leading behavior of the cut-off function is independent on initial conditions and for large $k$ 
the cut-off can be well approximated by a Gaussian form. Section IV is devoted to generalized 
attachment kernel $k+a_0$. Now both the exponent $\gamma\neq 3$ and the cut-off depend on $a_0$. 
We calculate the cut-off function explicitly and show that it is no longer Gaussian but exhibits 
the scaling predicted in \cite{kr2}. Section V deals with non-tree graphs. Some mathematical formulas 
we exploit in the paper are discussed in Appendices.

\section{Finite size effects in a growing scale-free tree}
For the presentation of our method we begin with the simplest variant of
the model, namely the BA tree network with $m=1,n_0=2$ and without initial attractiveness. 
The mathematical approach developed in this Section will be applied to more involved situations
in Sections III and IV.

We start from putting down the rate equation for $N_k(N)$, 
the average number of nodes with degree $k$ for the network consisting of $N$ nodes.
For simplicity we assume that the growth process starts from a dimer configuration, that is $n_0=2$ 
nodes joined by a link. Thus, $N_k(2)=2\delta_{k,1}$, where $\delta_{kq}$ stands for the Kronecker delta. 
At each time step a new node is added and connected to a node $i$ chosen at random out of 
$N$ old vertices with probability proportional to its degree $k_i$, thus obtaining a new network with 
$N+1$ nodes. The network is a tree graph. The recursion formula for $N_k(N)$ has the form
\ba
	& & N_k(N+1) = N_k(N) + \delta_{k,1}+\nonumber \\
	&+& \frac{k-1}{2(N-1)} N_{k-1}(N) - \frac{k}{2(N-1)} N_{k}(N).
	\label{rN}
\ea
The equation is exact for any $N$.
The Kronecker delta stands for the introduction of a new node with degree one. 
The next term corresponds to the appearance of a node with degree $k$ as a result of 
link's addition to the node with degree $k-1$. 
The last term describes how fast nodes with degree $k$  disappear due to link's addition to those nodes. 

In the limit $N\to\infty$ the solution is given by $N_k \approx N\pi^{BA}_k$ where, according to Ref. \cite{kr},
\bq
	\pi^{BA}_k = \frac{4}{k(k+1)(k+2)}. \label{piba}
\eq
In this paper we are interested in the general solution for $N_k(N)\equiv N \pi_k(N)$, with $\pi_k(N)$ 
being the degree distribution of a finite network. 
Our method to obtain this distribution is based on using its moments. 
We shall do this in several steps. First, we consider moments for $\pi_k(N)$, which are of particular importance for many problems including 
epidemic spreading and percolation \cite{ep}. Let $M_n(N)$ be the $n$-th moment of $N_k(N)$ and $\mu_n(N)$ be 
the $n$-th moment of the degree distribution $\pi_k(N)$:
\ba
	M_n(N) &=& \sum_{k=1}^\infty k^n N_k(N), \label{Mkdef} \\
	\mu_n(N) &=& \sum_{k=1}^\infty k^n \pi_k(N)  = \frac{1}{N} M_n(N).
\ea
Multiplying both sides of Eq. (\ref{rN}) by $k^n$ and summing over $k$ we get now recursive relations 
between moments $M_n(N)$ for different $N$. For instance, the equations for the first three moments read:
\ba
	M_0(N+1) &=& M_0(N) + 1, \label{M0} \\
	M_1(N+1) &=& \frac{2N-1}{2(N-1)} M_1(N) + 1, \label{M1} \\
	M_2(N+1) &=& \frac{2N}{2(N-1)} M_2(N) + \frac{M_1(N)}{2(N-1)}+1, \label{M2}
\ea
with the initial condition $M_n(2)=2$ imposed by Eq. (\ref{Mkdef}) for the dimer configuration.
The important feature of these (and higher) relations is that for any $n$, $M_n(N+1)$ depends only on lower moments. 
We can thus solve these equations for $M_n(N)$ by forward substitution, 
starting from the lowest moment. Equation (\ref{M0}) has an obvious solution $M_0(N) = N$. Solution of 
Eq. (\ref{M1}) is also easy to guess, being just twice the number of links: $M_1(N) = 2(N-1)$. In order 
to solve Eq. (\ref{M2}) we insert the solution for $M_1$ and we solve the resulting equation by standard 
methods. We get
\bq
	M_2(N) = 2(N-1) H(N-1),
\eq
where $H(n)=\sum_{i=1}^n 1/i$ is the harmonic number. Higher moments can be computed using similar tricks, but the formulas get much more complicated. The appropriate  moments for $\pi_k(N)$ are
\ba
	\mu_0 &=& 1, \\	
	\mu_1 &=& 2-2/N, \\
	\mu_2 &=& (2-2/N)H(N-1). \label{mu2} 
\ea
We see, that $\mu_2\approx 2\log N$ diverges for large $N$, but is still quite small for 
networks of order $10^9$ nodes, the largest real-world structures like the WWW. 
This suggests that real SF growing networks exhibit strong finite size
effects which is indeed true \cite{kr} and the assumption that $\mu_2=\infty$ for the BA model
is far from reality even for very large networks.

The equation for $\mu_2$ gives a rough estimate for the position of the cut-off in the degree
distribution as imposed by finite size effects.
Let us assume the cut-off in the degree distribution to be sharp enough and take $\pi_k$ to be 
$\pi_k=\pi^{BA}_k$ for $k$ smaller than $k_{max}$ and vanishing above $k_{max}$. 
Then the second moment behaves as $4\log k_{max}$. Comparing this with exact Eq. (\ref{mu2}) 
we get $k_{max} \approx \sqrt{N}$. The position of the cut-off scales therefore as a square root of 
the network size which in fact has been pointed out by several authors \cite{extr1,extr2}. 
We will show further explicitly that such scaling indeed holds for BA model without initial attractiveness. 

To find the correction to $\pi^{BA}_k$ in the finite network we introduce a function $v_k(N)$ defined 
as $v_k(N) \equiv N_k(N)/ \pi^{BA}_k$. The function $v_k(N)$ gives finite size corrections.
From the asymptotic properties of $\pi_k$ in the limit of large $N$ we know that $v_k(N)\propto N$ for $k\ll k_{max}$ and falls to zero far above the critical $k_{max}$. Equation (\ref{rN}) can be rewritten
in terms of $v_k(N)$:
\ba
	v_k(N) &=& \frac{3}{2} \delta_{k,1} + \frac{2+k}{2(N-2)} v_{k-1}(N-1) - \nonumber \\
	&-&  \frac{4-2N+k}{2(N-2)} v_k(N-1).
	\label{rv}
\ea
To proceed, we define moments $m_n(N)$ for the distribution $v_k(N)$:
\bq
	m_n(N) = \frac{1}{N-1} \sum_{k=1}^\infty k^n v_k(N),
\eq
with the value of the normalization constant $1/(N-1)$ motivated by later convenience.
Multiplying now both sides of Eq. (\ref{rv}) by $k^n$ and summing over $k=1,\dots,\infty$ we get
\ba
	m_n(N+1) &=&  \frac{1}{2N}\biggl( 3+ \sum_{i=0}^{n-1} c_{ni} m_i(N) + \biggr. \nonumber \\
	& + & \biggl. (2N+n+1)m_n(N) \biggr), 
        \label{rm}
\ea
where the initial condition
\bq
m_n(2) = 3	\label{ic1}
\eq
stems from the configuration of the starting graph and the coefficients $c_{ni}$ are given by
\bq
c_{n0} = 3,\;\; \mbox{and}\; \;\, c_{ni} = 3\binom{n}{i}+\binom{n}{i-1}\;\;  \mbox{for}\;\; i>0.
\label{Cik}
\eq
By solving equations for the first few moments we can infer the general solution:
\bq
	m_n(N) = \frac{1}{\Gamma(N)} \sum_{i=0}^{n+1} \frac{B_{ni}}{\Gamma(2+i/2)} \Gamma(N+i/2), \label{gsm}
\eq
where the coefficients $B_{ni}$ still have to be found and the factor $\Gamma(2+i/2)$ is singled out 
for convenience. Inserting this to Eq. (\ref{rm}) we get after some manipulations:
\ba
	\sum_{i=0}^{n+1} B_{ni} (i-n-1) \frac{\Gamma(N+i/2)}{\Gamma(2+i/2)} =
	 3\Gamma(N)+ \nonumber \\ +\sum_{j=0}^{n-1} c_{nj} B_{j0} \Gamma(N)
	+ \sum_{i=1}^n \sum_{j=i-1}^{n-1} c_{nj} B_{ji} \frac{\Gamma(N+i/2)}{\Gamma(2+i/2)}.
\ea
Comparing terms of the same order in $\Gamma$-functions we get two recursion relations for $B_{ni}$:
\bq
	B_{n0} = -\frac{1}{n+1} \left( 3+\sum_{j=0}^{n-1} c_{nj} B_{j0} \right) \label{ak0}
\eq
for $i=0$ and
\bq
	B_{ni} = -\frac{1}{n+1-i} \sum_{j=i-1}^{n-1} c_{nj} B_{ji} \label{aki}
\eq
for $0<i\leq n$. The third recursion relation which completes the set comes from the initial condition 
(\ref{ic1}):
\bq
	B_{n,n+1} = 3 - \sum_{i=0}^n B_{ni},
	\label{akk1}
\eq
and $B_{00}=-3$. Returning to $m_n(N)$ we find that for large $N$ the leading term reads:
\bq
	m_n(N) \simeq B_{n,n+1} \frac{\Gamma\left(N+\frac{n+1}{2}\right)}{\Gamma(N)\Gamma\left(2+\frac{n+1}{2}\right)} 
        \simeq N^\frac{n+1}{2} A_n,
\eq
with $A_n = B_{n,n+1}/\Gamma\left(\frac{5+n}{2}\right)$. Such scaling indicates that for large $N$,
\bq
	v_k(N) \simeq N w(k/\sqrt{N}), \label{wq}
\eq
where $w(x)$ is a function having $A_n$ as its  moments:
\bq
	A_n = \int_0^\infty dx\, w(x) x^n.	\label{akw}
\eq
Therefore the degree distribution for large but finite BA tree network is well approximated by
\bq
	\pi_k(N) = \pi^{BA}_k w\left(k/\sqrt{N}\right).
\eq
The functional form $w(x)$ of the cut-off can be found analytically by reconstructing it from its moments $A_n$, which express through the coefficients $B_{n,n+1}$. 
However, in order to calculate them one has to know all $B_{ni}$ resulting from
Eqs. (\ref{ak0}),(\ref{aki}) and (\ref{akk1}). Fortunately, they can be conveniently stored by means 
of appropriately defined generating functions, for which we can write differential equations and solve 
them. In Appendix A we show that
\bq
	A_n = \frac{(2+n)^2 n!}{\Gamma((3+n)/2)}.	\label{akfinal}
\eq
The behavior of the cut-off function $w(x)$ for large values of the argument is dominated by the behavior of $A_n$ for large $n$ which can be easily obtained using Stirling's formula for the factorial and for the Euler's Gamma function:
\bq
	\log A_n \approx \frac{1}{2} n \log n.	\label{aklarge}
\eq
Let us now compare it with moments $I_n$ of the function $\exp\left[ -(x/\sigma)^\alpha\right]$:
\bq
	I_n = \int_0^\infty x^n \exp\left[ -(x/\sigma)^\alpha\right] dx = \frac{\sigma^{n+1}}{\alpha} \Gamma
	\left( \frac{n+1}{\alpha} \right).	\label{Ik}
\eq
For large $n$ the leading term of $\log I_n \approx (n \log n)/\alpha$ is the same as in Eq. (\ref{aklarge})
with $\alpha=2$, which means that the tail of $w(x)$ falls like a Gaussian function. 
The parameter $\sigma$ can be established by comparing sub-leading terms in $I_n$ and $A_n$.
The value $\sigma=2$ obtained in this way will be confirmed below by direct calculation of $w(x)$.

From the values of the coefficients $A_n$, the function $w(x)$ can be restored using
a trick involving inverse Laplace transform. In Appendix B we show how to get $w(x)$ expressed as an infinite series:
\bq
	w(x) = 1-\frac{4}{\sqrt\pi} \sum_{n=1}^\infty x^{2n+1} \frac{(-1)^n n^2}{n!2^{2n}(2n+1)}.
	\label{finalwx}
\eq
One can check that this result corresponds to a Taylor series for an expression involving the complementary error function $\mbox{erfc}(z)$:
\bq
	w(x) = \mbox{erfc}(x/2)+\frac{2x+x^3}{\sqrt{4\pi}} e^{-x^2/4},
\eq
the result given in \cite{kr2}, 
which is close to an approximate result of \cite{dr1}.
The series in Eq. (\ref{finalwx}) is rapidly convergent and, if truncated at some $n_{max}$, 
can serve for numerical calculations.

\section{Arbitrary initial condition}
In this section we generalize the method to the case when the network's
initial configuration is a complete graph with more than two nodes. We proceed exactly as before.
First we write the rate equation for the average number of nodes of a given degree.
Second, we factorize the solution into two terms: the one corresponding to the thermodynamical limit 
$N\to\infty$ and another one $v_k(N)$ giving the correction for finite $N$. Then we find  moments 
for $v_k(N)$ and deduce the form of $w(x)$ from their asymptotic properties using formulas 
given in Appendices. Although we still add one link per one added node, we shall consider here a more general rate equation for $N_k(N)$:
\ba
	N_k(N+1) &=& N_k(N) + \delta_{k,m} +\frac{k-1}{2(N-\omega)} N_{k-1}(N) - \nonumber \\
	&-& \frac{k}{2(N-\omega)} N_{k}(N), \label{rNm0}
\ea
with two free parameters $m$ and $\omega$. The parameter $m$ corresponds to the number
of links added with one new node. In the cases considered in the previous and in the 
present section we have $m=1$. The case $m>1$ is considered in Sec. \ref{mlarge}.  
The parameter $\omega$ in the denominators in Eq. (\ref{rNm0}) is introduced to guarantee 
the normalization of the probability of preferential attachment and depends on the form of 
the preferential attachment kernel and on initial conditions.
For the linear attachment kernel $k$ the normalization condition is $\sum k N_k(N)=2L$, with
$L$ being the overall number of bonds. We define $\omega$ to fulfill the equation: $2L=2(N-\omega)$.
If we start with a complete graph with $n_0$ nodes then $\omega =n_0(3-n_0)/2$, which in particular 
gives $2L=2(N-1)$ for the dimer-configuration from Sec. II.
Here we first consider Eq. (\ref{rNm0}) for arbitrary $m$ and $\omega$ and put appropriate 
values of these parameters at the end of computations.

\begin{figure}
\psfrag{x}{$x$} \psfrag{wx}{$w(x)$}
\includegraphics[width=8cm]{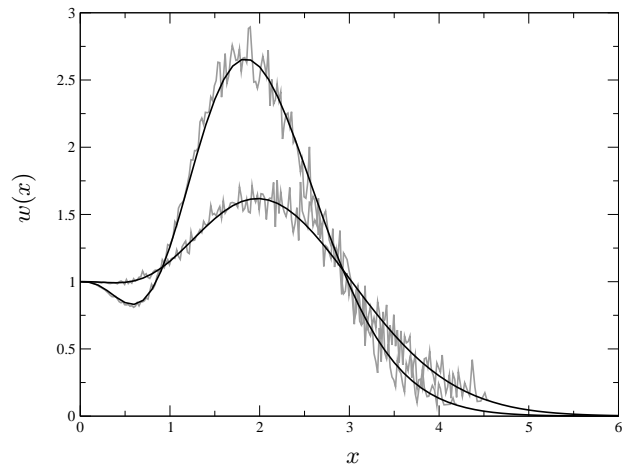}
\caption{The cut-off function $w(x)$ giving finite-size corrections to the degree distribution, 
calculated from Eq. (\ref{wxm0}) for the starting graph with $n_0=3$ (black lower line) and $5$ nodes 
(black upper line). Our analytical results agree very well with $w(x)$ obtained from averaged degree 
distributions for $2\times 10^4$ generated networks of size $N=10^4$ (gray lines).
}
\label{bagen2}
\end{figure}

In the thermodynamical limit the degree distribution 
is given by \cite{kr}:
\bq
  \pi_{\infty}(k) = \frac{2m(m+1)}{k(k+1)(k+2)},	\label{pimm0}
\eq
for all $k\geq m$ and does not depend on $\omega$. 
Like in the previous section, we insert $N_k(N) = \pi_k^{BA} v_k(N)$ and get the equation for $v_k(N)$. 
Then, defining the moments $m_n(N)\equiv \sum_k k^n v_k(N)/(N-\omega)$ we obtain the recursion relation:
\ba
	m_n(N+1) &=& \frac{\scriptstyle 1}{\scriptstyle 2(N+1-\omega)}\biggl( (2+m)m^n + \sum_{i=0}^{n-1} c_{ni} m_i(N) + \biggr.\nonumber \\
	&+& \biggl. (2N-2\omega+3+n)m_n(N) \biggr), \label{rmm0} \\
	m_n(n_0) &=& \frac{(n_0^2-1)n_0^2}{2m(m+1)(n_0-\omega)} (n_0-1)^n ,	\label{ic1m0}
\ea
where the initial condition (\ref{ic1m0}) has now a more complicated form than before. 
The general solution for $m_n(N)$ can still be written by means of Gamma functions:
\bq
	m_n(N) = \frac{1}{\Gamma(N+1-\omega)} \sum_{i=0}^{n+1} B_{ni}\frac{\Gamma(N+1-\omega+i/2)}{\Gamma(n_0+1-\omega+i/2)}
	 , \label{gsmm0}
\eq
which imposes that $v_k(N)$ scales as a function of $x\equiv k/\sqrt{N}$ exactly like in Eq. (\ref{wq}).
The equations for $B_{ni}$ look similar to Eqs. (\ref{ak0}), (\ref{aki}) and (\ref{akk1}) but with
the constant 3 changed for $(2+m)m^n \Gamma(n_0+1-\omega)$ in the first and for $m_n(n_0)\Gamma(n_0+1-\omega)$
in the third equation. Proceeding as in Section II we find generating functions for 
$B_{ni}$ and finally, the expression for $A_n$:
\ba
	A_n &=& \frac{\Gamma(1+n_0-\omega)}{\Gamma(n_0+3/2-\omega+n/2)} \biggl[
	\frac{\Gamma(3+m+n)}{(n+1)\Gamma(m+2)} + \biggr. \nonumber \\
	&+& \biggl. \frac{m_0(n_0) \Gamma(2+n_0+n)}{\Gamma(n_0+2)} \biggr].
	\label{akfinalm0}
\ea
The generating function $M(z)$ will therefore be a sum of two $f$-functions defined in Eq. (\ref{MZgen}),
see Appendix B. Using the results of Appendix C we can immediately write the expression for the cut-off 
function $w(x)$:
\ba
	w(x) &=& \Gamma(1+n_0-\omega) \biggl[ \frac{1}{\Gamma(m+2)} \tilde{f}_{\frac{1}{2},n_0+\frac{3}{2}-\omega,
	3+m,2}(x) + \biggr. \nonumber \\ 
	&+& \biggl. \frac{m_0(n_0)}{\Gamma(n_0+2)} \tilde{f}_{\frac{1}{2},n_0+\frac{3}{2}-\omega,
	2+n_0,1}(x) \biggr],	\label{wxm0}
\ea
where $\tilde{f}$-functions are defined in Appendix C and $\omega$, $m_0$ have been given above and depend only on $n_0$. 
In figure \ref{bagen2} we plot $w(x)$ calculated from Eq. (\ref{wxm0}) together with
the results of numerical simulation for finite size networks. One readily infers a strong dependence of 
$w(x)$ on the size of the seed graph $n_0$. This sensitivity to the initial conditions has just 
been reported in \cite{kr2} as well as in Ref. \cite{malarz} where another quantity was measured.
However, if one compares the asymptotic behavior of $A_n$'s:
\bq
	\log A_n \approx \frac{1}{2} n \log n - \frac{n}{2}\left(1-\log 2\right),
\eq
with that of Eq. (\ref{Ik}), one immediately finds that for large $n$ the function $w(x)$
still behaves like $\exp(-x^2/4)$, independently of $n_0$. Therefore the degree distribution for the BA tree model
without initial attractiveness has always a Gaussian cut-off whose position scales as $\sim N^{1/2}$.

\section{Generalized attachment kernel}
The considerations above can be easily extended to the case of preferential attachment kernel $k+a_0$ 
with initial attractiveness $a_0>-1$. In Ref.~\cite{kr} it is shown that the degree distribution in the 
thermodynamical limit reads:
\bq
	\pi^\infty_k = \frac{(2+a_0)\Gamma(3+2a_0)}{\Gamma(1+a_0)}\frac{\Gamma(k+a_0)}{\Gamma(k+3+2a_0)}.
\eq
For large $k$, $\pi^\infty_k$ decays according to a power-law $\sim k^{-\gamma}$ with $\gamma=3+a_0$.
Therefore this model can reproduce any exponent $\gamma$ observed in real-world  networks.
In \cite{kr} it has been shown that the model is equivalent to the growing network with re-direction 
(GNR), which is a convenient method for simulating the model on a computer. At each step we choose 
at random one node $i$ from the set of existing nodes. A newly introduced node is attached with 
probability $1-r$ to this $i$, and with probability $r$ to its ancestor, i.e. to the node the proper, 
oriented link of $i$ was attached to when this node was itself introduced into the network. With the 
choice $r=1/(a_0+2)$ the GNR model is equivalent to the BA tree model with initial attractiveness. 
In all numerical simulations showed in this section the GNR model is used.
A generalized version of the GNR model which works for $m>1$ links added per one new node is described in \cite{abrah}.

The recursion formula for $N_k(N)$ is slightly modified in comparison to Eq. (\ref{rN}) and reads:
\ba
	N_k(N+1) &=& N_k(N)  +\frac{k+a_0-1}{N(2+a_0)-2} N_{k-1}(N) - \nonumber \\
	&-& \frac{k+a_0}{N(2+a_0)-2} N_{k}(N) + \delta_{k,1}, \label{rNgen}
\ea
but the initial conditions remain the same as in Section II. 
Inserting now $N_k(N)= v_k(N) \pi^\infty_k$ we get the equation 
for the function $v_k(N)$ and then the equation for the moments $m_n(N)$:
\ba
	m_n(N+1) = \frac{\scriptstyle 3+2a_0}{\scriptstyle N(2+a_0)} + \frac{\scriptstyle N-1}{\scriptstyle \big((N-1)(2+a_0)+a_0\big)N} \times \nonumber \\
	\times \left[ \sum_{i=0}^{n-1} c_{ni} m_i(N) + \big(N(2+a_0)+a_0+1+n\big)m_n(N)\right], \nonumber \\
	\label{rmgen}
\ea
where $c_{ni}$'s read now:
\ba
&&  c_{n0} = 3+2a_0, \nonumber \\
&& c_{ni} = (3+2a_0)\binom{n}{i}+\binom{n}{i-1}, \quad \mbox{for } i \ge 1.
\ea
Following the same steps as in Section II we can find the general functional form of $m_n(N)$ and its leading behavior for large networks:
\ba
	m_n(N) &\simeq &  \frac{B_{n,n+1}\Gamma\left(N+\frac{a_0+n+1}{2+a_0}\right)}{\Gamma\left(N-1+\frac{a_0}{2+a_0}
	\right)\Gamma\left(2+\frac{a_0+n+1}{2+a_0}\right)(N-1)} \nonumber \\
	&\simeq & N^\frac{n+1}{2+a_0} A_n,
\ea
with $A_n=B_{n,n+1}/\Gamma\left[2+\frac{a_0+n+1}{2+a_0}\right]$. Therefore the function $v_k(N)$ obeys the following 
scaling rule:
\bq
	v_k(N) \to N w\left(k/N^\frac{1}{2+a_0}\right), \label{wqgen}
\eq
where the function $w(x)$ has  moments $A_n$ depending on $a_0$. 
Equation (\ref{wqgen}) indicates that the cut-off scales as $N^{1/(\gamma-1)}$ where $\gamma=3+a_0$ is the 
exponent in the power-law for $\pi^\infty_k$. This is in agreement with arguments given in \cite{kr2}.
For a given size $N$, the cut-off shifts to lower values of $N$ when the exponent $\gamma$ increases. 
This implies that the power-law in the degree distribution can hardly be seen for 
$\gamma>4$, because even for large networks with $N=10^6$ nodes
the cut-off corresponds to the value of $k_{max}\sim 100$ and the power-law lasts only for $1-2$ decades 
in $k$. This partially explains the fact that the SF networks with $\gamma$ above $4$ are practically 
unknown \cite{cn2}.

\begin{figure}
\psfrag{x}{$x$} \psfrag{wx}{$w(x)$}
\includegraphics[width=8cm]{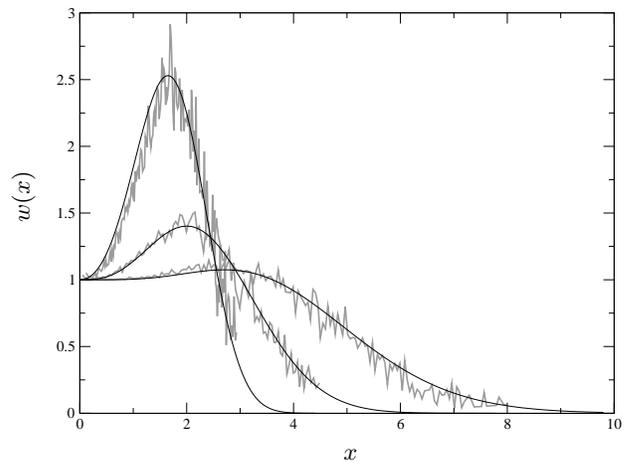}
\caption{Plots of $w(x)$ calculated from Eqs. (\ref{coeff}) and (\ref{finalwxgen})
for $a_0=-0.5$, $0$ and $1$ (solid lines from top to bottom) which correspond to $\gamma=2.5,3$
and $4$, respectively. The curves become flat with increasing $a_0$. The thick gray lines are 
$w(x)$ obtained from averaged degree distributions for $2\times 10^4$ networks of size $N=10^4$.
The tails decay like $\exp(-x^\alpha)$ with $\alpha=3,2,3/2$, respectively, in agreement with numerical findings.}
\label{bagen1}
\end{figure}

As before, we can find a general formula for $A_n$ and estimate its leading behavior:
\bq
	\log A_n \approx \frac{1+a_0}{2+a_0} n \log n.
\eq
Comparing this to Eq. (\ref{Ik}) as it has been done before, one obtains 
that for large $x$ the function $w(x)$ decays like $\exp\left[ -(x/\sigma)^\alpha\right]$ 
with
\bq
\alpha=\frac{2+a_0}{1+a_0}=\frac{\gamma-1}{\gamma-2}.
\eq
This agrees very well with numerical findings and means that the cut-off 
for $\gamma\neq 3$ is no longer of a Gaussian type. For $2<\gamma<4$, as often found in real networks,
$\alpha$ is always larger than $1.5$ and therefore the cut-off $w(x)$ cannot be approximated by a 
pure exponential decay observed in some networks \cite{cn2}; exponential cut-offs most likely have 
a different origin than the finite size effects. The value of $\sigma$ can also be obtained from 
subleading terms and is given by
\bq
	\sigma = (2 + a_0) (1+a_0)^{-1/\alpha}.
\eq
The result for $M(z)$ has now the form of Eq. (\ref{MZgen}) with 
\ba
	&\mathcal{N}&=\frac{2\Gamma(1+\frac{a_0}{2+a_0})}{\Gamma(5+2a_0)}, \;
	a=6+3a_0, \; b=12+13a_0+4a_0^2, \nonumber \\ 
	&\alpha &=\frac{1}{2+a_0}, \; \beta=\frac{5+3a_0}{2+a_0}, \; 
	\xi=4+2a_0, \; \zeta=2,	\label{coeff}
\ea
and therefore $w(x)$ is given by Eq. (\ref{finalwxgen}) with the corresponding values of parameters. 
In Fig. \ref{bagen1} we plot $w(x)$ for $a_0=-1/2,\,0$ and $1$. For the purpose of numerical calculations 
all series have been truncated, with the error being less than $10^{-4}$ in the plotted area. 
These results show that the curves become more flat with increasing $a_0$ and 
agree well with $w(x)$ obtained in simulations of finite size networks.

As before, the starting graph has large influence on the exact form of $w(x)$. 
We do not consider here the dependence on the size $n_0$ of the seed graph, however, just like in 
Sec. III, one can show that asymptotic properties of the cut-off function are insensitive to $n_0$
and therefore for $x$ being sufficiently large, $w(x)\sim \exp\left[ -(x/\sigma)^\alpha\right]$ depends only on $a_0$ i.e. only on the exponent $\gamma$ in the power-law $\pi_k\sim k^{-\gamma}$.

\section{BA model with $m>1$ - beyond the tree model}
\label{mlarge}
In the previous sections we have considered growing networks restricting ourselves to the case when graphs
are essentially trees (the possible cycles can only stem from the seed graph).
The general case of $m>1$ links joined to a new node in a single step is much more complicated. 
Each of $m$ proper links of a newly introduced node has to be connected to one of $N$ preexisting 
nodes according to the preferential attachment rule. However, since multiple links are not allowed, 
the nodes to which links have been connected on this step have to be excluded from the set of nodes 
available for further linking. Thus, when a new link is introduced, the probabilities of attaching it to
one of the preexisting nodes depend whether the link is the first, second, ..., etc. of $m$. 
The rate equation for $N_k(N)$ can still be obtained in this case. However, its structure is very 
sophisticated and it is highly non-linear in $N_k$ and $k$. This makes impossible 
the immediate application of our method. To proceed, we note that for large $N$ and $m\ll N$ the 
probability of choosing the same node as a candidate for link's attachment more than once in a 
single time step is very small. We can thus write the rate equation fully disregarding the exclusion of 
multiple connections, since their relative number should be small in the thermodynamic limit 
$N\to\infty$. For instance, for $m=2$ the approximate equation for $N_k(N)$ reads:
\ba
	N_k(N+1) &=& N_k(N) + \delta_{k,2} + \frac{k-1}{2N-3}N_{k-1}(N) - \nonumber \\
	&-&\frac{k}{2N-3}N_{k}(N).
	\label{rNm=2approx}
\ea
We assume additionally $N_k(3)=3\delta_{k,2}$ that is we start from a complete graph with $n_0=3$ 
(triangle). In the general case $m\geq 2$ the approximate equation takes the form of Eq. (\ref{rNm0}). 
The origin of all terms is the following. The Kronecker delta gives the addition of a node with $m$ 
links at each time step. The number of links $L=m(N-\omega)$ with 
\bq
	\omega=n_0(2m+1-n_0)/(2m)	\label{omm0}
\eq
gives the normalization factor $2L$ for the  probability of attachment.
The factor $m$ hidden in $2L$ cancels out with $m$ possibilities of choosing links at each step.
For $m=2$ and $n_0=3$ it reduces to Eq. (\ref{rNm=2approx}).

Like before, we expect some dependence on starting graph, but as long as asymptotic properties of $w(x)$
are concerned, its particular choice does not seem to be important.
It is therefore tempting to assume $n_0=m+1$ because it allows to have $\pi_k=0$ for all $k<m$ at each 
time step. However, we must remember that the Eq. (\ref{rNm0}) with $\omega$ given by Eq. (\ref{omm0}) 
is a reasonable approximation of the unknown true rate equation
for $N_k(N)$ only if $m \ll N$ at each stage of the network's growth. Thus our expressions for $w(x)$ 
works well only for $m\ll n_0$.

We can check the validity of Eq. (\ref{rNm0}) calculating
the moments $\mu_n(N)$ of the degree distribution $\pi_k$.
The equation (\ref{rNm0}) has a form similar to Eq. (\ref{rN}) and thus all results are similar to those presented in Sections II-IV. However the approximation we made includes only leading terms in the limit of large $N$
and the quality of that approximation can only be checked by computer simulations. 
For instance, for $n_0=m+1$ the first three moments $\mu_n(N)$ are
\ba
	\mu_0 &=& 1, \\
	\mu_1 &=& m(2-(m+1)/N)=2L/N , \\
	\mu_2 &=& \left(2-\frac{\scriptstyle m+1}{\scriptstyle N}\right) \left( m^2+\sum_{i=m+1}^{N-1} \frac{m(m+1)}{2i-m+1} \right).
\ea 
First two moments are exact although we have used the approximate equation.
For large $N$ the last expression reduces to
\ba
	\mu_2 &\approx & \left(2-\frac{\scriptstyle m+1}{\scriptstyle N}\right)\biggl[ m^2+\frac{m(m+1)}{2} \biggl( \gamma-\log 2- \biggr.\biggr. \nonumber \\
	\biggl.\biggl. &-& H\left(\frac{\scriptstyle 1+m}{\scriptstyle 2}
		\right)+\log(2N-m-1) \biggr) \biggr], \label{mu2gen}
\ea
where $H(n)$ are harmonic numbers and $\gamma\approx 0.5772$ is the Euler-Mascheroni constant.
Thus the second moment grows like $m(m+1)\log N$ and similarly to what has been done in Sec. II one can infer
that the cut-off scales like $\sqrt{N}$.
In table \ref{tab1} we compare the quality of formula (\ref{mu2gen}) for $m=2$ with $\mu_2$ found for 
numerical solutions of exact equation for $N_k(N)$. The good agreement confirms that the approximation we have made 
performs well.
\begin{table}
$$ 
\begin{array}{|c|c|c|c|}
\hline
N & \mu_2 \; \mbox{exact} & \mu_2 \; \mbox{from Eq. (\ref{mu2gen})} & \mbox{error in} \% \\
\hline
200 & 34.16 & 35.26 & 3.2 \\
\hline
1000 &  43.73 & 45.15 & 3.2 \\
\hline
2000 &  47.83 & 49.34 & 3.1 \\
\hline
\end{array}
$$
\caption{Comparison between exact and approximate value of the second moment $\mu_2$ in case of $m=2$.}
\label{tab1}
\end{table}

\begin{figure}
\psfrag{x}{$x$} \psfrag{wx}{$w(x)$}
\includegraphics[width=8cm]{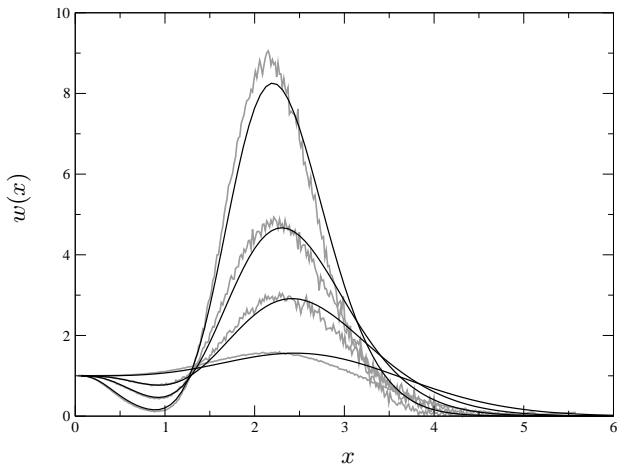}
\caption{The function $w(x)$ for $m=2$ and for $n_0=3,7,10$ and $15$ (curves from the flattest to the most peaked).
The agreement between the analytical solution and simulation results is not as good as before 
due to approximate character of solved equation. The simulation data were obtained for $N=4000$. 
The dip at about $x=1$ and the peak at about $x=2$ especially pronounced in the case of larger 
seed graphs mean that for large starting graph much more nodes with high degrees is present than
it would be expected for the asymptotic power-law behavior of $\pi^\infty_k$.
}
\label{bagen3}
\end{figure}

Let us go now to the corrections to the degree distribution. In the thermodynamic limit,
$\pi^\infty_k$ is given by Eq. (\ref{pimm0}). The degree distribution for large but finite networks 
corresponds to $\pi^\infty_k$  multiplied by $w(k/\sqrt{N})$. From Sec. III we know that $w(x)$ is expressed
by Eq. (\ref{wxm0}) where $m$ and $n_0$ are now completely arbitrary and $\omega$ is given by 
Eq. (\ref{omm0}). The asymptotic form of $w(x)\sim \exp(-x^2/4)$ resulting from 
analytical formula, is the same as in Sec. III.
In figure \ref{bagen3} we compare our approximate analytical solution with results of direct numerical
simulations for different $n_0$. A small deviation between analytical and numerical curves is evident. 
This deviation is the largest for $n_0=3$ and the smallest for $n_0=15$, confirming that the performance
of our approximation is the better, the larger is the seed graph. 

\section{Conclusions}
We have investigated finite size corrections to the degree distribution $\pi_k$
in the Barab\'{a}si-Albert model of growing network, assuming preferential attachment kernel $k+a_0$.
Imposing that $\pi_k$ is a product of known distribution $\pi^\infty_k\sim k^{-\gamma}$ for infinite 
network and the function $w(x)$ giving the finite size cut-off in the rescaled variable 
$x= k/N^\frac{1}{\gamma-1}$, we have reduced the problem to finding the moments of $w(x)$. 
We have shown that for sufficiently large $N$ they grow faster than exponentially and give the 
explicit formula for their leading behavior. We have found the expression for $w(x)$ and argued that far 
above the cut-off it behaves as $\sim \exp(-(x/\sigma)^\alpha)$ with $\alpha=(\gamma-1)/(\gamma-2)$ for 
any exponent $\gamma$ lying in the range $2 <\gamma < \infty$. 
We discuss the dependence of $w(x)$ on the initial configuration of the network and show that although 
it has strong influence on exact form of $w(x)$, its asymptotic properties are independent on the seed graph.
Moreover, we give a general expression for $w(x)$ in terms of convergent power series in $x$ and
present plots for some special cases.

The approach we have developed here allows for treating the general BA model
in a new fashion. This approach can be also applied to other 
models of growing networks provided the recursion equations for the connectivity $N_k(N)$ 
are known. 


\section*{Acknowledgments}
We thank G. Pagnini for providing us with important information on Mittag-Leffler and H Fox functions,
and Z. Burda and S. N. Dorogovtsev for valuable discussions.
B. W. was supported by the EU grant MTKD-CT-2004-517186 (COCOS) and by F. Kogutowska Foundation for 
Jagellonian University.

\appendix
\section{}

Let us define the following generating functions for $B_{ni}$:
\bq
	F_0(x) = \sum_{n=2}^\infty B_{n0} \frac{x^n}{n!} ,\label{F0}
\eq
and
\bq
	F_i(x) = \sum_{n=i}^\infty B_{ni} \frac{x^n}{n!} ,\label{Fi}
\eq
for $i>0$. Note that Eq. (\ref{F0}) is not a special case of Eq. (\ref{Fi}).
We start from Eq. (\ref{ak0}), multiplying its both sides by $(n+1)x^n/n!$ and summing over $n=2,\dots,\infty$.
As a result one can easily show that $F_0(x)$ satisfies the differential equation:
\bq
	(e^x-1)F_0'(x) = (2-3e^x)F_0(x) - 3(3x-1)(e^x-1)-3x,
\label{EqA3}
\eq
with $F_0(0)=0$. Equation (\ref{EqA3}) has the following solution:
\bq
	F_0(x) = 2-3x-e^{-2x}-e^{-x}. \label{f0x}
\eq
In a similar way, from Eq. (\ref{aki}) and using the definition of $c_{ni}$ from Eq. (\ref{Cik}), 
we get the differential equation for $F_i(x)$:
\ba
	(1-e^x)F_i'(x) &=& (3e^x-2-i)F_i(x) + B_{i-1,i} \frac{x^{i-1}}{(i-1)!}\times \nonumber \\
	 &\times &\left[ 3e^x-3+\frac{i-1}{x}(e^x-1-x) \right].
\ea
This complicated equation has a simple solution due to the fact that $F_i(x)$ contains only powers of $x$ higher than $i-1$:
\bq
	F_i(x) = B_{i-1,i} \frac{e^{-(2+i)x} (e^x-1)^{i-1} - x^{i-1}}{(i-1)!}. \label{fix}
\eq
Now, if we multiply Eq.~(\ref{akk1}) by $x^n/n!$ and sum over $n=1,\dots,\infty$, we find:
\bq
	\sum_{n=1}^\infty \frac{x^n}{n!} B_{n,n+1} + \sum_{i=0}^\infty F_i(x) = 3(e^x-x-1).
	\label{g1}
\eq
Substituting $F_i(x)$ by Eqs.~(\ref{f0x}) and (\ref{fix}) and defining 
a new generating function for the coefficients $B_{n,n+1}$:
\bq
	G(z) = \sum_{n=1}^\infty \frac{z^n}{n!} B_{n,n+1},
\eq
we get the expression:
\bq
	e^{-3x}G(1-e^{-x}) = 3e^x+1+e^{-2x}+e^{-x}-6e^{-3x},
\eq
as follows from Eq. (\ref{g1}). This equation for $G(z)$ can be readily solved:
\bq
	G(z) = 3(1-z)^{-4}+(1-z)^{-3}+(1-z)^{-2}+(1-z)^{-1}-6.
	\label{Gz}
\eq
The values of $B_{n,n+1}$ can be obtained  by using Cauchy's formula as integrals over a contour encircling the point $z=0$ in the complex plane:
\bq
	B_{n,n+1} = \frac{n!}{2\pi i} \oint \frac{dz}{z^{n+1}} G(z) = n!\frac{(2+n)^2 (3+n)}{2}.
\eq
Finally, from the definition of $A_n = B_{n,n+1}/\Gamma\left(\frac{5+n}{2}\right)$ we get Eq. (\ref{akfinal}).

\section{}
To get the functional form of the cut-off $w(x)$ we define a generating function:
\bq
	M(z) = \sum_{n=0}^\infty A_n \frac{z^n}{n!}.
\eq
Comparing this definition with Eq. (\ref{akw}) we see that $M(z) = \int_0^\infty \exp(zx) w(x) dx$ so that 
\bq
	M(-z) = \int_0^\infty \exp(-zx) w(x)\, dx
\eq
is the Laplace transform of $w(x)$. Therefore $w(x)$ is given by the inverse Laplace transform of $M(z)$
or equivalently by the Fourier transform of $M(-iz)$:
\bq
	w(x)	= \frac{1}{2\pi} \int_{-\infty}^{\infty} dz\, e^{izx} M(-iz).	\label{invlap}
\eq
Using the explicit form of coefficients $A_n$ we get
\bq
        M(z)= \sum_{n=0}^\infty \frac{(2+n)\Gamma(n+3)}{\Gamma(n+2)\Gamma((3+n)/2)} z^n.
	\label{Mzf}
\eq
This series has an infinite radius of convergence. 
The function $M(z)$ given by Eq. (\ref{Mzf}) is a special case of more general power series:
\bq
	M(z) = \mathcal{N} \sum_{n=0}^\infty \frac{(an+b)\Gamma(n+\xi)}
        {\Gamma(n+\zeta)\Gamma(\alpha n+\beta)} z^n,
	\label{MZgen}
\eq
which depends on six parameters $a,b,\alpha,\beta,\xi,\zeta$ and where $\mathcal{N}$ gives the appropriate normalization.
We introduce an auxiliary function $f_{\alpha,\beta,\xi,\zeta}(z)$:
\bq
  f_{\alpha,\beta,\xi,\zeta}(z) = 
  \sum_{n=0}^\infty \frac{\Gamma(n+\xi)}{\Gamma(n+\zeta)\Gamma(\alpha n+\beta)}z^n .
\eq
This one is a special case of a Fox-Wright Psi function \cite{moskowitz} or of a Fox H function 
(see e.g. \cite{pagnini}). Its Fourier transform can be written as follows:
\bq
	\tilde{f}_{\alpha,\beta,\xi,\zeta}(x) = \frac{1}{2\pi} \int_{-\infty}^{\infty} dz\, e^{izx} f(-iz)
	= I_C(x)-I_S(x),
\eq
where $I_C(x)$ and $I_S(x)$ are cosine and sine transforms of real and imaginary part of $f(-iz)$, respectively:
\ba
	I_S(x) = \frac{1}{\pi} \int_0^\infty \sin(zx) \Im f(-iz) dz, \\
	I_C(x) = \frac{1}{\pi} \int_0^\infty \cos(zx) \Re f(-iz) dz.
\ea
We consider here only $I_C(x)$ since repeating  all calculations presented below for $I_S(x)$ 
one can check that \mbox{$I_C(x)=-I_S(x)$}.
The function $I_C(x)$ can be expressed via the Mellin transform $f^*_C(s)$ of $\Re f(-iz)$:
\bq
	I_C(x) = \frac{1}{\pi x} \frac{1}{2\pi i} \int_{-i\infty}^{i\infty} f^*_C(s) \Gamma(1-s)\sin(\pi s/2) x^s ds,
\eq
where
\bq
	f^*_C(s) = \int_0^\infty z^{s-1} \Re f(-iz) dz.
\eq
The real part of $f(-iz)$ reads:
\bq
  \Re f(-iz) = \sum_{n=0}^\infty \frac{(-z^2)^n \Gamma(2n+\xi)}{\Gamma(2n+\zeta)\Gamma(2\alpha n+\beta)}.
\eq
We now substitute two of three Gamma functions by their integral representations:
\ba
	\Gamma(2n+\xi) &=& \int_0^{\infty} e^{-t} t^{2n+\xi-1} dt, \label{intgamma1} \\
	\frac{1}{\Gamma(2n+\zeta)} &=& \int_{\rm H} e^u u^{-2n-\zeta} du, \label{intgamma2}
\ea
where ``H'' denotes the Hankel contour. Changing variables $z\to v: z=v \frac{u}{t}$, we get the 
following integral representation:
\ba
	f^*_C(s) = \int_0^\infty dt \, e^{-t} t^{\xi-1-s} \int_{Ha} du \, e^u u^{s-\zeta} \times \nonumber \\
	\times \int_0^\infty dv \, v^{s-1} \sum_{n=0}^\infty \frac{(-v^2)^n}{\Gamma(2\alpha n+\beta)}.
\ea
The last integral is now just the Mellin transform of the Mittag-Leffler function defined as
\bq
	E_{2\alpha,\beta}(t) = \sum_{n=0}^\infty \frac{t^n}{\Gamma(2\alpha n+\beta)}. \label{MLfunct2}
\eq
From \cite{marichev} p. 301 we have
\bq
	E_{2\alpha,\beta}(-v^2) \stackrel{\rm Mellin}{\longleftrightarrow} \frac{\Gamma(s/2)\Gamma(1-s/2)}{2\Gamma(\beta-\alpha s)},
\eq
and after simple algebraic manipulations we finally arrive at:
\bq
	f^*_C(s) = \frac{\Gamma(\xi-s)\Gamma(s/2)\Gamma(1-s/2)}{2\Gamma(\zeta-s)\Gamma(\beta-\alpha s)}.
\eq
Now we are able to calculate $\tilde{f}(x)=2I_C(x)$ as an inverse Mellin transform.
Using the identity for Gamma functions
\bq
	\Gamma(t)\Gamma(1-t) = \frac{\pi}{\sin(\pi t)}, \label{gammy}
\eq
and applying the residue theorem we get:
\ba
	& & \tilde{f}_{\alpha,\beta,\xi,\zeta}(x) = \frac{1}{2\pi i} \int_{-i\infty}^{i\infty} ds\, x^{s-1}
	\frac{\Gamma(\xi-s)\Gamma(1-s)}{\Gamma(\beta-\alpha s)\Gamma(\zeta-s)} \nonumber \\
	&=& \sum_n \mbox{res}_{s_n} \left[ \frac{\Gamma(\xi-s)\Gamma(1-s)}{\Gamma(\zeta-s)\Gamma(\beta-\alpha s)} x^{s-1}
	 \right]_{s=s_n}, \label{fftgen}
\ea
where the sum runs over all points $s_n$ at which either $\Gamma(1-s)$ or $\Gamma(\xi-s)$ has a pole.
The above formula simplifies for $\xi,\zeta$ being positive integers $m,k$:
\bq
	\tilde{f}_{\alpha,\beta,m,k}(x) = \sum_{n=0}^\infty (-x)^n \frac{\scriptstyle(m-2-n)(m-3-n)\cdots (k-1-n)}
{\scriptstyle\Gamma(\beta-\alpha-\alpha n)n!}, \label{fftsimpl}
\eq
where we make use of the fact that $\mbox{res}_{z=-n}\, \Gamma(z) = (-1)^n/n!$.
Let us come back to $M(z)$ which can be expressed through $f(z)$ and its derivative $f'(z)$:
\bq
	M(z) = \mathcal{N} \left( azf'_{\alpha,\beta,\xi,\zeta}(z)+bf_{\alpha,\beta,\xi,\zeta}(z)\right).
\eq
Then $w(x)$ is given by Fourier transforms of the two functions $f(z)$ and $zf'(z)$. Integrating by parts 
we can change from $zf'_{\alpha,\beta,\xi,\zeta}(z)$ to
 $f_{\alpha,\beta-\alpha,\xi-1,\zeta-1}(z)+\mbox{const}$, where the constant term
vanishes under the transform, so that we finally get:
\bq
	w(x) = \mathcal{N} \left( ax \tilde{f}_{\alpha,\beta-\alpha,\xi-1,\zeta-1}(x)+
	(b-a)\tilde{f}_{\alpha,\beta,\xi,\zeta}(x)\right), \label{finalwxgen}
\eq
where $\tilde{f}$'s are given either by Eq. (\ref{fftsimpl}) or by more general Eq. (\ref{fftgen}).

Applying now Eq. (\ref{finalwxgen}) to $\mathcal{N}=1,a=1,b=2,\alpha=1/2,\beta=3/2,\xi=3,\zeta=2$ as stems from Eq. (\ref{Mzf}), we get the function $w(x)$ for the BA tree model:
\ba
	w(x) &=& x\tilde{f}_{1/2,1,2,1}(x)+\tilde{f}_{1/2,3/2,3,2}(x) = \nonumber \\
	&=& \sum_{n=0}^\infty \frac{(-x)^n}{n!}
	\left[ \frac{-nx}{\Gamma(1/2-n/2)}+\frac{1-n}{\Gamma(1-n/2)} \right]. \nonumber \\
\ea
With help of Eq. (\ref{gammy}) 
we get after some algebraic manipulations:
\bq
	w(x) = 1-\frac{4}{\pi} \sum_{n=1}^\infty x^{2n+1} (-1)^n n^2 \frac{\Gamma(n+1/2)}{\Gamma(2n+2)}.
\eq
Finally, having in mind that
\bq
	\Gamma(2n+2) = \frac{2^{2n}}{\sqrt\pi} n!(2n+1)\Gamma(n+1/2),
\eq
we get Eq. (\ref{finalwx}).

\end{document}